\documentclass[final]{elsart}
\newcommand{\beq}{\begin{equation}}
\newcommand{\eeq}{\end{equation}}
\usepackage{graphicx}
\usepackage{amssymb}
\journal{Physics Letters B}
\begin{document}

\begin{frontmatter}

\title{Systematic thermal reduction of neutronization in core-collapse supernovae}

\author[MI,Orsay]{A. F. Fantina},
\author[MI]{P. Donati},
\author[MI]{P. M. Pizzochero\corauthref{cor}}
\corauth[cor]{Corresponding author.}
\ead{pierre.pizzochero@mi.infn.it}

\address[MI]{Dipartimento di Fisica, Universit\`a degli
Studi di Milano, and Istituto Nazionale di Fisica Nucleare,
sezione di Milano, Via Celoria 16, 20133 Milano, Italy}

\address[Orsay]{Institut de Physique Nucl\'eaire, Universit\'e Paris-Sud,\\
IN2P3-CNRS, 91406 Orsay Cedex, France }

%\date{\today}

\begin{abstract}

We investigate to what extent the temperature dependence of the
nuclear symmetry energy can affect the neutronization of the
stellar core prior to neutrino trapping during gravitational
collapse. To this end, we implement a one-zone simulation to
follow the collapse until $\beta$-equilibrium is reached and the
lepton fraction remains constant. Since the strength of electron
capture on the neutron-rich nuclei associated to the supernova
scenario is still an open issue, we keep it as a free parameter.
We find that the temperature dependence of the symmetry energy
consistently yields a small reduction of deleptonization, which
corresponds to a systematic effect on the shock wave energetics:
the gain in dissociation energy of the shock has a small yet
non-negligible value of about 0.4 foe (1 foe $ = 10^{51}$ erg) and
this result is almost independent from the strength of nuclear
electron capture. The presence of such a systematic effect and its
robustness under changes of the parameters of the one-zone model
are significant  enough  to justify further investigations with
detailed numerical simulations of supernova explosions.

\end{abstract}
\begin{keyword}
supernova  \sep gravitational collapse \sep neutronization \sep
symmetry energy
 \PACS 97.60.Bw \sep  26.50.+x \sep 23.40.-s
\end{keyword}

\end{frontmatter}

\section{\label{sec1}Introduction}

 Weak interaction processes are naturally associated to
 core-collapse supernovae and more generally to compact stars.
 Indeed, on the one hand the increasing density in the collapsing
 core of a massive star continuously shifts the $\beta$-equilibrium
 conditions and thus drives electron capture (first mostly on
 exotic nuclei and then on free, unbound protons) all the way to an almost
 completely deleptonized equilibrium state, the final neutron star. On the
 other hand, the tremendous densities and temperatures obtained through the
 gravitational compression allow the neutrinos, produced both thermally and in
 such weak processes, to interact significantly with matter. The neutrinos
 diffuse, rather than stream freely, out of the collapsing core
 and such neutrino transport produces unique physical scenarios, like neutrino
 trapping half-way in the collapse  or the shock-wave revival
 powered by neutrino cooling of the proto-neutron star. We refer
 to the review by Hans Bethe \cite{Be} for a masterful physical
 discussion of these phenomena and of their relevance to supernova
 explosions.

 In a previous paper \cite{DPBB} (from now on Paper~I; see also Ref.~\cite{Erice}
for further details), we discussed a particular issue related to
electron capture in collapsing stellar cores. First, we studied
the temperature dependence of the nucleon effective mass,
$m^{\star}$, in the nuclei $^{98}$Mo, $^{64}$Zn and $^{64}$Ni as
due to the coupling of the mean field single-particle levels to
the collective surface vibrations of the nucleus, calculated in
the quasi-particle random phase approximation (QRPA). Then, we
observed that the decrease with temperature obtained for
$m^{\star}$ in the range $0 < T < 2 $ MeV induces a corresponding
increase of the nuclear symmetry energy (per nucleon), $E_{\rm
sym}$, and in analogy to the results of the Fermi gas model we
argued that the temperature dependence of $E_{\rm sym}$ can be
fitted by a simple analytical expression. Finally, we investigated
the implications of such a temperature dependence on the
gravitational collapse of the core of massive stars. We did this
in a one-zone (uniform mean density) model, an approach which
incorporates the important physics but is easy to implement and
which, in the past, has proven effective to make a preliminary
study of the core deleptonization during the infall epoch before
core bounce, when the collapse is still homologous
\cite{EP,Fu,RCK}. In Paper~I, electron capture was implemented as
in the classic work by Bethe \emph{et al.} (BBAL) \cite{BBAL}, but
with the strength of capture on nuclei quenched by a factor
$\gamma^2 = 0.1 $ to account for the Pauli blocking of
Gamow-Teller  (GT) transitions in neutron-rich heavy nuclei
discussed in Refs.~\cite{Fu,CW}. The collapse simulation showed
that the temperature dependence of the symmetry energy yields a
lower rate of neutronization along the collapse, as expected for
larger values of $E_{\rm sym}$, and thence a higher value for the
electron fraction, $Y_e$, at neutrino trapping density. This can
be conveniently quantified in terms of the associated \emph{gain}
in dissociation energy of the shock, $\delta_{_T} E_{\rm diss}$, a
quantity which gives a more direct physical insight\footnote{A
larger lepton fraction after trapping corresponds to a larger
homologous core so that the shock wave, which forms at its edge
after core bounce, will have less material to traverse before
getting out of the iron core and thence it will dissipate less
energy in the photo-dissociation of tightly bound nuclei
\cite{Be}.}. The results of Paper~I correspond to an energy gain
$\delta_{_T} E_{\rm diss} \sim 0.5-0.6$ foe (1 foe $= 10^{51}$
erg), a non-negligible amount when one considers that the
explosion energy (kinetic energy of the ejecta) of SN 1987A was
observed to be $\sim 1$ foe \cite{BP}.

After Paper~I was published, two more investigations of the
temperature dependence of the nuclear symmetry energy, due to the
presence of neutron-rich heavy nuclei in the collapsing core, have
been presented \cite{DKLR,DLS}; both are based on shell model
Monte Carlo (SMMC) calculations, the model of choice to take into
account nuclear correlations beyond those treated at the QRPA
level. More recently, the virial EOS of hot ($T \gtrsim 2$ MeV)
low-density nuclear matter composed of protons, neutrons and alpha
particles was obtained \cite{HS}; such a scenario is
representative of matter around the neutrinosphere after core
bounce and shock formation, during the contraction phase of the
proto-neutron star and the associated loss of gravitational energy
through neutrino emission. The results found in Ref.~\cite{HS} for
the temperature dependence of the symmetry energy follow from the
presence of these alpha clusters. Such findings, however, cannot
be applied to the pre-bounce scenario studied here; indeed, the
temperature of the infalling matter reaches 2~MeV only long after
$\beta$-equilibrium is achieved and the composition is always
dominated by neutron-rich heavy clusters in a sea of unbound
neutrons, while the fraction of alphas is negligible at these low
temperatures. The $T$-dependence of $E_{\rm sym}$, derived in
Refs.~\cite{DPBB,DKLR,DLS} and under discussion here, follows from
the presence of these exotic nuclear species.

In Ref.~\cite{DKLR}, several isobaric pairs with mass numbers in
the range $ A = 54-64$ were studied. Although the results obtained
for the nuclei $^{64}$Zn and $^{64}$Ni were in agreement with
those of Paper~I, in their conclusions the authors claimed to
''find no systematic temperature dependence of the symmetry energy
coefficient\footnote{The coefficient $b_{\rm sym}(T)$ is the same
as the coefficient $s(T)$ of Paper~I, and it is related to $E_{\rm
sym}(T)$ by the standard expression $E_{\rm sym} = b_{\rm sym} (1-
2x)^2 $, with $x=Z/A$ for a nucleus of mass $A$ and charge $Z$.},
$b_{\rm sym}$, for $T \leq 1$ MeV. This contradicts a recent
suggestion that $b_{\rm sym}$ increases by 2.5 MeV at this
temperature'' \cite{DKLR}. An improved SMMC calculation, however,
was presented several year later in Ref.~\cite{DLS}, where some
known problems of the previous paper (small model space and
g-extrapolation procedure to circumvent the notorious sign problem
of SMMC) had been fixed. Nine isobaric pairs with $A= 56-66$ were
analyzed and this time the authors concluded that their ''SMMC
studies are consistent with an increase of the symmetry energy
with temperature, supporting the argumentation of Donati \emph{et
al.}'' \cite{DLS}. Indeed, upon averaging over the various pairs,
they found  a variation $\overline{\delta b}_{\rm sym} =(6.2 \pm
1.8) \%$ in the temperature interval  $T= 0.33-1.23$ MeV, which is
in reasonable agreement with the QRPA results of Paper~I, namely
an increase of the symmetry energy of $\sim 8 \%$ in the interval
$T= 0 -1$ MeV.

In the concluding section of Ref.~\cite{DLS}, the authors also
quickly discussed possible consequences for core-collapse
supernovae. They studied the decrease of electron capture on
nuclei due to the proposed temperature dependence of the symmetry
energy, by considering the increase of reaction Q-values induced
by it. For the neutron-rich nuclei expected during collapse (mass
number $A > 65 $), they adopted new capture rates obtained in the
so-called ''hybrid'' model \cite{LKD}, an approximate approach
which mixes SMMC and RPA techniques to go beyond the independent
particle model (IPM) in the calculation of both allowed and
forbidden transitions. These new results show unblocking of the GT
strength \cite{LKD}, due to configuration mixing by the residual
interaction and to thermal excitations, which significantly modify
the naive single-particle occupations of the IPM and thus yield
capture rates one order of magnitude larger than those predicted
by the IPM \cite{Fu}. Proceeding in this way, the authors obtained
changes of electron capture rates due to the $T$-dependence of
$b_{\rm sym}$ that ''appear to be rather mild so that one does not
expect significant changes for the collapse trajectory''
\cite{DLS}. Although we agree that no dramatic effect on the
dynamics of the collapse is to be expected, one should be more
cautious in dismissing any significant consequence of the
$T$-dependence of $E_{\rm sym}$ without a collapse simulation.
Indeed, not only the reaction Q-values (as considered in
Ref.~\cite{DLS}), but also the equation of state of bulk dense
matter, the free nucleon abundances, the degree of dissociation
into $\alpha$-particles and the nuclear internal excitations are
affected by changes in the symmetry energy. Moreover, the dynamics
of the collapse depends in a very non-linear way on the strength
of nuclear electron capture\footnote{The parameter study of
Ref.~\cite{Hal2}, for example, shows that each increase of the
rate of capture by a factor 10 corresponds roughly to the same
decrease ($\sim 0.1 M_{\odot}$) of the mass of the homologous
core.}, so that mild changes in the rates may still result in
non-negligible alterations of the overall energetics.

The purpose of the present article is to investigate with a
collapse simulation the extent to which the temperature dependence
of the nuclear symmetry energy, found in Paper~I and confirmed in
Ref.~\cite{DLS}, can affect the deleptonization of the collapsing
stellar core. We must, of course, take into account the remarkable
progress made in the SMMC calculations of electron capture rates
since publication of Paper~I. On the one hand, the new values
obtained with improved SMMC techniques for capture on nuclei
present in lower-density matter ($A < 65$) \cite{LM} have been
implemented in modern evolutionary stellar calculations yielding
new presupernova models \cite{HLMW}, which are significantly
different than those used so far as initial conditions in collapse
simulations. On the other hand, the unblocked GT strengths found
with the hybrid model for the neutron-rich nuclei typical of
higher-density matter ($A>65$) \cite{LKD} have been used in
numerical (1-dimension) collapse simulations, both newtonian and
relativistic. When compared to the results from the commonly used
Bruenn parametrization of nuclear electron capture \cite{Br},
which quenches capture on heavy nuclei as required by the IPM and
thus allows capture on free (unbound) protons to dominate some
crucial phases of the collapse, the simulations with the new rates
show significant differences in the dynamics of the shock wave and
in the neutrino luminosity \cite{Lal,Hal1,Hal2}.

Altogether, the results obtained in the collapse simulations of
Paper~I have to be revisited in  four main aspects, all related to
electron capture on nuclei:
\\
\emph{i)}  by using the approach of BBAL \cite{BBAL} in Paper~I,
we certainly overestimated the effect of the temperature-dependent
symmetry energy on the deleptonization. Indeed, the BBAL rates for
capture on nuclei are calculated applying the Fermi approximation
to a shell model description of the GT transition. This
statistical limit (which actually does not apply to the collapse
scenario, where the shell structure is still dominant and the
nuclear density of states is far from thermal
 \cite{LKD}) involves an integration over the
initial proton states and this multiplies the final capture rates
by a factor containing the nucleon effective mass\footnote{The
integration requires the nuclear density of states, which in the
Fermi gas model is proportional to the nucleon mass \cite{BBAL}.}.
This linear dependence of the nuclear rates on $m^{\star}$
obviously amplifies the thermal effects, but it is absent if a
more realistic, non-statistical description of capture is adopted.
\\
\emph{ii)} the BBAL rates for electron capture (on both nuclei and
free protons) used in Paper~I were calculated at $T=0$, but since
we are looking for a small thermal effect we cannot neglect the
influence of the Fermi distribution functions,  which describe  the
occupation numbers of initial and final particle states at
 finite temperature \cite{EP,Fu}.
\\
\emph{iii)} the multiplying factor $\gamma^2 = 0.1 $, introduced
in Paper~I to account for the Pauli blocking of GT transitions, is
not anymore realistic according to the new results from the hybrid
model \cite{LKD}. These new findings, however, are not yet
obtained in a consistent SMMC calculation so that, in our opinion,
the actual strength of nuclear electron capture is still an open
issue and the correct value of $\gamma^2$ is not yet pinned down.
\\
\emph{iv)} the initial conditions  adopted in Paper~I for the
collapse have to be revisited, to account for the new results
obtained for the presupernova core when implementing the improved
SMMC capture rates in evolutionary stellar codes \cite{HLMW}.

In the next section, we describe our model for the gravitational
collapse of the stellar core and discuss how it takes into proper
account all these issues.

\section{\label{sec2}Physical model for the collapse}

In order to study the neutronization of matter induced by
gravitational collapse, we develop a one-zone model (sphere of
uniform density) along the classic approach of
Refs.~\cite{EP,Fu,RCK}. The model is an improvement over the one
used in Paper~I in two respects: first, the treatment of electron
capture is revisited in order to answer the issues \emph{i)} and
\emph{ii)} previously mentioned; then, the trapping of neutrinos
is treated more realistically and provides the equilibrium lepton
fraction after trapping, when the collapse is adiabatic. Moreover,
the capture strength on nuclei is kept as a free parameter,
$\gamma^2$, as discussed in issue \emph{iii)}, and the
presupernova initial conditions are the improved ones mentioned in
issue \emph{iv)}.

We now  describe the main features of our collapse model:
\\
 \emph{(1)} The dynamical
evolution of density with time due to gravity decouples from the
thermodynamical equations for the changes in entropy and lepton
fractions. Therefore, we can follow the relevant thermodynamical
variables (entropy, temperature, electron and neutrino fractions,
particle abundances, nuclear composition) as a function of
density, along the so called collapse trajectories.
\\
\emph{(2)} We adopt the equation of state (EOS) for hot dense
matter derived in BBAL \cite{BBAL}.  The ensemble of nuclear
species is approximated by a mean heavy nucleus\footnote{An
ensemble of nuclei is actually present, in nuclear statistical
equilibrium under strong and electromagnetic interactions. The
mean nucleus is the one that minimizes the nuclear energy and thus
it represents the most abundant nuclear species \cite{Fu}.} in a
sea of dripped-out free neutrons and (fewer) protons. The
fractions of free nucleons are determined from nuclear statistical
equilibrium. The symmetry energy appears in the bulk nuclear
energy and, as a consequence, in the neutron chemical potential,
$\mu_n$, and in the neutron-proton energy difference, $\hat{\mu}=
\mu_n - \mu_p $. These are crucial quantities in determining the
free particle abundances, the nuclear capture Q-values and the
entropy changes due the departure from $\beta$-equilibrium of the
collapsing core before neutrino trapping.
\\
 \emph{(3)} Thermal dissociation of nuclei into
$\alpha$-particles and nucleons is also taken into account through
the Saha equation, but found to have a negligible effect on the
collapse trajectories.
\\
\emph{(4)} Entropy terms are included for the translational
degrees of freedom of all the particles (mean heavy nucleus, free
classical nucleons, relativistic degenerate leptons) as well as
for the internal nuclear excitations, treated in the Fermi gas
approximation . The nuclear excitation energy is proportional to
the nucleon effective mass (see Ref.~\cite{Be}), which is the
quantity whose temperature dependence we originally calculated and
fitted by an analytical expression in Paper~I. We find that the
corresponding entropy term has a non-negligible effect on the
collapse trajectories.
\\
\emph{(5)} Neutrino trapping is set to start at a given trapping
density, $\rho_{\rm{tr}}$. The typical ''standard'' value is
$\rho_{\rm{tr},10}= 43$ ($\rho_{10}$ being the density in units of
$10^{10}$ g cm$^{-3}$), but we keep it as a model parameter. As
long as $\rho < \rho_{\rm{tr}}$, neutrinos are allowed to stream
freely out of the core and the neutrino fraction is $Y_{\nu}= 0$.
When $\rho \geq \rho_{\rm{tr}}$, neutrino diffusion is treated
along the lines of Ref.~\cite{RCK}: a degenerate sea of neutrinos
with $Y_{\nu}\neq 0$ is allowed to build up by the inclusion of a
diffusion term which decreases with density. Moreover, the inverse
reactions induced by the sea of neutrinos are included in the
electron capture rates \cite{LV}, so that weak interactions can
reach equilibrium. In this way, complete neutrino trapping is
reached gradually at a density somewhat larger than
$\rho_{\rm{tr}} \, $; both the total lepton fraction, $Y_{l}=
Y_{e} + Y_{\nu}$, and the entropy tend naturally to constant
values, after which the collapse proceeds adiabatically and in
$\beta$-equilibrium. This is a major improvement over Paper~I,
where neutrinos were always streaming out freely ($Y_{\nu}= 0$),
so that equilibrium could never be reached and the equilibrium
lepton fraction was just the value of the electron fraction taken
at $\rho=\rho_{\rm{tr}}$ along the collapse trajectory, namely $
Y_e = Y_e(\rho_{\rm{tr}})$.
\\
\emph{(6)} Electron capture is implemented on both free protons
and heavy nuclei with standard two-level transitions, as fully
developed in Ref.~\cite{Fu}; the phase space integral is
calculated numerically, although its approximation by Fermi
integral (as in Eq.~(1) of Ref.~\cite{Lal}) turns out to be
accurate enough. For this kind of transitions, the nuclear capture
rate $\lambda_{\rm N}$ is a function of density, temperature and
two other quantities: the excitation energy of the nuclear GT
resonance, $ \Delta_{\rm N}$, and the reaction Q-value. The first
is taken as a model parameter, while $ Q = \hat{\mu} + \Delta_{\rm
N} $ (we have actually used a regularized expression for the GT
excitation energy \cite{LV}). We have also multiplied the nuclear
strength $\lambda_{\rm N}$ by a free parameter, $\gamma^2$. As
shown in Ref.~\cite{Lal}, the Q-dependence of the capture rates
obtained with the hybrid model can be reasonably fitted by the
two-level expression, with $\Delta_{\rm N} = 2.5$ MeV and an
appropriate GT matrix element; we normalize $\lambda_{\rm N}$ so
that our expression coincides with Eq.~(1) of Ref.~\cite{Lal} when
$\gamma^2 = 1$.
\\
\emph{(7)} The temperature dependence of the symmetry energy  is
treated as in Paper~I \cite{DPBB,Erice}, where it was expressed in
terms of the  $T$-dependence of the nucleon effective mass,
$m^{\star} = m^{\star}(T)$, calculated for different nuclei. The
results for each nucleus were fitted with a formula containing two
parameters: the value at $T=0$ of the so-called $\omega$-mass, $
m_{\omega}(0)$, and the temperature scale of this dependence,
$T_0$. The standard average values are $ m_{\omega}(0)= 1.7 $ and
$ T_0 = 2 $ MeV, but we keep them as model parameters allowed to
vary in a meaningful physical range ($1.4 \lesssim m_{\omega}(0)
\lesssim 1.8$ and $ 1.9 \lesssim T_0 \lesssim 2.1$ MeV
\cite{DPBB}), to account for their dependence on the nucleus
studied.

Details and equations can be found in
Refs.~\cite{DPBB,Erice,EP,Fu,RCK,BBAL}. In particular, we adopted
the following  differential equations for the collapse
trajectories: \\
 \emph{i)} the electron fraction evolution, $\mathrm{d} Y_e / \mathrm{d} \rho
 $, is Eq.~(85) of Fuller \cite{Fu}. From trapping density
 onwards, the neutrino-induced inverse reactions are included as in
 Eqs.~(15) and (16) of Ray \emph{et al.} \cite{RCK}.
 \\
 \emph{ii)} the entropy and temperature evolution,
 $\mathrm{d} S / \mathrm{d} \rho $ and $\mathrm{d} T / \mathrm{d} \rho
 $, are respectively: Eqs.~(92) and (93) of Fuller  before trapping\footnote{In Eq.~(92) of Fuller, the terms $A X_p$
 in the denominators should each be multiplied by  a factor $\lambda_{\mathrm{fp}}$.}; Eqs.~(33) and (35) of Ray \emph{et
 al.} after trapping.
\\
 \emph{iii)} the neutrino fraction evolution, $\mathrm{d} Y_\nu / \mathrm{d} \rho
 $, is Eq.~(29) of \emph{Ray et al.} after trapping; before trapping we take $Y_\nu = 0$.

The collapse trajectories are determined starting from a set of
initial conditions on the density, $\rho_{\rm i}$, the
temperature, $T_{\rm i}$, and the electron fraction, $Y_{e,{\rm
i}} $ (until trapping density is reached, $Y_{\nu}=0$). According
to the improved results of Ref.~\cite{HLMW} for the central
properties of the presupernova core which evolves from a $ 15
M_{\odot}$ star (about the size of the progenitor of SN 1987A), we
will take the initial values $\rho_{10,{\rm i}} = 0.936$, $T_{\rm
i} = 0.625$ MeV and $Y_{e,{\rm i}}=0.432 $, which differ
significantly from those adopted in Paper~I. The differential
equations are then integrated and the collapse trajectories of the
different thermodynamical quantities are found. In particular, the
total lepton fraction $Y_{l} = Y_{l}(\rho)$ tends to a constant
value, $ Y_{l,{\rm tr}}$, as the density increases above
$\rho_{\rm{tr}}$ and neutrino trapping is completed.

In the next section, we discuss our results for the neutronization
of the core in terms of  $ Y_{l,{\rm tr}}$ and of quantities
related to it.

\section{\label{sec3}Results of the collapse simulation}

 We first fix the model parameters to their ''standard'' values
 ($\rho_{\rm{tr},10} = 43 $, $ m_{\omega}(0)= 1.7 $, $ T_0 = 2 $
MeV, $ \Delta_{\rm N}  = 2.5 $ MeV) and make a parameter study of
the core neutronization as a function of the nuclear strength in
the range $0 \leq \gamma^2 \leq 5$. We point out that $\gamma^2=0$
corresponds to electron capture on free protons \emph{only}, while
$\gamma^2=5$ is very large and probably unrealistic. The older,
blocked GT rates of Fuller \cite{Fu} correspond to $\gamma^2=0.1$,
while the new unblocked rates of Ref.~\cite{Lal} are associated to
 $\gamma^2=1$. Improved future calculations could change
the presently accepted value of the nuclear strength, but (barring
discovery of past errors or unexpected breakthroughs) we think
that  $0.5 \lesssim \gamma^2 \lesssim 2$ should represent a
reasonable physical range.

For each choice of parameters, we have run the collapse simulation
twice: once implementing the temperature dependence  $E_{\rm
sym}=E_{\rm sym}(T)$  and obtaining $ Y_{l,{\rm tr}}|_{_T}$, once
setting $E_{\rm sym}=E_{\rm sym}(0)$ and obtaining $ Y_{l,{\rm
tr}}|_{_0}$. We indicate by $ \delta_{_T}$ the ''thermal''
variation of a quantity \emph{due} to the temperature dependence
of the symmetry energy; for example, the thermal change in
equilibrium lepton fraction is $ \delta_{_T} Y_{l,{\rm tr}} =
Y_{l,{\rm tr}}|_{_T} - Y_{l,{\rm tr}}|_{_0}$.

\begin{figure}[!ht]
\begin{center}
\includegraphics[width=10cm]{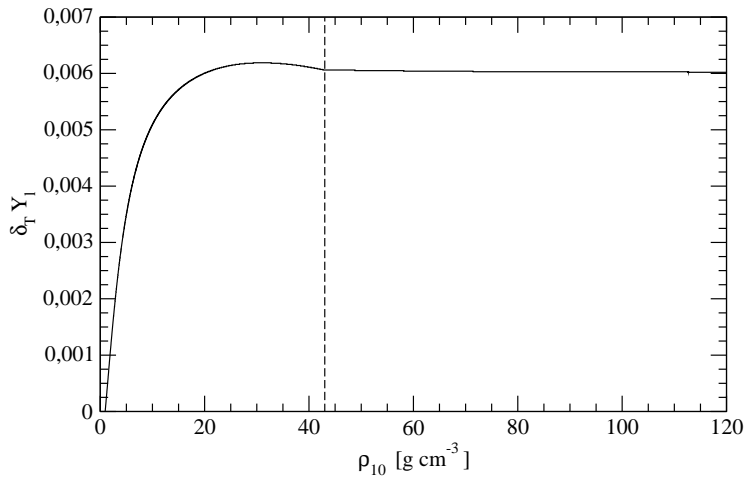}
\end{center}
\caption{Thermal variation of the lepton fraction collapse
trajectories, $ \delta_{_T} Y_{l}$, as a function of the density
of the collapsing core, $\rho_{10}$  (in units of $10^{10}$ g
cm$^{-3}$). The dotted line indicates the trapping density,
$\rho_{\rm{tr}}$. The curve corresponds to standard parameters of
the model ($\rho_{\rm{tr},10} = 43 $, $ m_{\omega}(0)= 1.7 $, $
T_0 = 2 $ MeV, $ \Delta_{\rm N} = 2.5 $ MeV).   } \label{fig1}
\end{figure}

From the collapse trajectories of the lepton fraction, $Y_{l} =
Y_{l}(\rho)$, we can infer the density range where the temperature
dependence of the symmetry energy is most relevant. Regarding the
thermal variation of the capture rates, we obtain similar results
as those shown in Figure~7 of Ref.~\cite{DLS}: the thermal effect
is maximum at the onset of collapse ($\rho_{10} \sim 1$) and
decreases with increasing density; when trapping sets in
($\rho_{10} \sim 40$), the difference in the capture rates due to
$E_{\rm sym}(T)$ is already negligible. However, since this is a
cumulative and non-linear effect on the lepton fraction, a
difference between the collapse trajectories, $ \delta_{_T}
Y_{l}(\rho) = Y_l(\rho)|_{_T} - Y_l(\rho)|_{_0} \, $, gradually
builds up as the density increases. This is shown in
Figure~\ref{fig1}, where the thermal variation of the lepton
fraction collapse trajectories, $ \delta_{_T} Y_{l} $, is given as
a function of the density of the collapsing core. We notice that,
even though the thermal variation actually reaches its final
equilibrium value, $ \delta_{_T} Y_{l,{\rm tr}}$, only  when
neutrino trapping is fully achieved (around $\rho_{10} \sim 250$),
the whole effect is seen to build up before trapping sets in. In
particular,  it is mostly in the low-density range $1\lesssim
\rho_{10}\lesssim 20$ that the temperature dependence of the
symmetry energy affects significantly the neutronization process.

Surprisingly, although Figure~\ref{fig1} corresponds to standard
parameters of the model, the curve $ \delta_{_T} Y_{l}(\rho) $ is
found to change very little under physically reasonable variations
of the model parameters, as will be discussed later on.

The results for the equilibrium lepton fractions and the
associated thermal changes are shown in Table~\ref{tab1} for
different values of $\gamma^2$. We point out how the general
magnitude of the equilibrium lepton fraction is a very slowly
decreasing function of $\gamma^2$. Increasing the strength by a
factor ten, from the blocked to the unblocked capture rates,
decreases the equilibrium lepton fraction by $ \sim 13\%$, which
is in reasonable agreement with the $\sim 10\%$ change obtained in
newtonian one-dimensional simulations \cite{Hal2}. We also notice
that the thermal effect under study \emph{systematically} reduces
the equilibrium neutronization, namely  $Y_{l,{\rm tr}}$ is
increased by an almost constant value, $\delta_{_T} Y_{l,{\rm tr}}
\simeq 0.006$, irrespective of the value of the strength
$\gamma^2$. Although small, this effect is not negligible, as we
will argue in the remaining of this article.

\begin{table}[!ht]
\begin{center}
\begin{tabular}{ccccc}
\hline\hline \ \ $\gamma^2 $ \ \ \ & \ \ $ Y_{l,{\rm tr}}|_{_0}$ \
\ & \ \ $ Y_{l,{\rm tr}}|_{_T}$ \ \ &
 $ \delta_{_T} Y_{l,{\rm tr}}$ &  $ \delta_{_T} E_{\rm diss}$  (foe)\\
\hline
0   & 0.3996 & 0.4054 & 0.0058 &  0.45  \\
0.1 & 0.3802 & 0.3861 & 0.0060 &  0.44  \\
0.5   & 0.3460 & 0.3519 & 0.0059 &  0.41 \\
1   & 0.3291 & 0.3351 & 0.0060 &  0.39 \\
2   & 0.3114 & 0.3175 & 0.0061 &  0.38 \\
5   & 0.2808 & 0.2868 &0.0060  &  0.34  \\
 \hline\hline
\end{tabular}
\end{center}
\caption{Results of the collapse simulation for different values
of the strength of nuclear electron capture, $\gamma^2$. We show
the equilibrium lepton fractions after trapping obtained without,
$ Y_{l,{\rm tr}}|_{_0}$,  and with, $ Y_{l,{\rm tr}}|_{_T}$, the
temperature dependence of the symmetry energy; the thermal change
of the equilibrium lepton fraction, $ \delta_{_T} Y_{l,{\rm tr}}
$; the corresponding gain in dissociation energy of the shock, $
\delta_{_T} E_{\rm diss}$ (in foe). The model parameters are the
standard ones ($\rho_{\rm{tr},10} = 43 $, $ m_{\omega}(0)= 1.7 $,
$ T_0 = 2 $ MeV, $ \Delta_{\rm N}  = 2.5 $ MeV).}

 \label{tab1}
\end{table}

In order to determine the relevance of our results to supernova
explosions, we need a quantity with a more direct physical meaning
and which can be compared to relevant observables. As in Paper~I,
we use the gain in shock dissociation energy which is defined as
 $ \delta_{_T} E_{\rm diss}=  98 \, [(Y_{l,{\rm tr}}|_{_T})^2 -
(Y_{l,{\rm tr}}|_{_0})^2] = 98 \ \delta_{_T} Y_{l,{\rm tr}}^2 $
{\rm (in foe). Although based on a schematic model for the shock
formation and propagation \cite{BBB}, this expression provides a
reasonable order of magnitude estimate of $ \delta_{_T} E_{\rm
diss}$. In a similar fashion, one could consider the change in
initial (i.e. post-bounce) shock energy, $ \delta_{_T} E_{\rm
shock}$, which also follows from changes in the equilibrium lepton
fractions affecting the size of the homologous core. In the
schematic approach of Ref.~\cite{Fu}, however, the expression for
the initial shock energy, $E_{\rm shock}=E_{\rm shock}(Y_{l,{\rm
tr}})$,  has a maximum for $Y_{l,{\rm tr}} = \frac{10}{13} Y_{i} =
0.3323$. Since the equilibrium lepton fractions corresponding to
$\gamma^2=1$ are close to this extremum (cf. Table~\ref{tab1}),
the thermal effect $ \delta_{_T} E_{\rm shock}$ turns out to be
quite small ($\sim 10^{-2}$ foe); we will not consider it in the
following.

Since $\delta_{_T} Y_{l,{\rm tr}}$ is small, the thermal gain in
dissociation energy can be written as $ \delta_{_T} E_{\rm diss}
\simeq 196 \ Y_{l,{\rm tr}}|_{_0} \times \delta_{_T} Y_{l,{\rm
tr}}$. This shows that in general $\delta_{_T} E_{\rm diss}$
depends on $\delta_{_T} Y_{l,{\rm tr}}$, but its magnitude is
fixed  by  the final neutronization reached by matter, $Y_{l,{\rm
tr}}|_{_0}$, which is determined by the nuclear capture strength
$\gamma^2$. In the last column of Table~\ref{tab1}, we show the
results for the gain in dissociation energy.  For standard
parameters and $\gamma^2 =1$, we find $\delta_{_T} E_{\rm diss}=
0.39$ foe. Moreover, since $\delta_{_T} Y_{l,{\rm tr}}$ is
constant, the gain in dissociation energy has the same very slow
dependence on the strength parameter as the equilibrium lepton
fraction. This is well seen in Figure~\ref{fig2}, where
$\delta_{_T} E_{\rm diss}$ is given as a function of $\gamma^2$.
The points are the results of the collapse simulation, while the
line in the log-log graph represent a power-law best fit, with a
very small exponent $m= -0.065$. In the physical meaningful range
for the strength ($0.5 \lesssim \gamma^2 \lesssim 2$), the gain in
dissociation energy  varies only by $\pm 4\% $, in the interval
$\delta_{_T} E_{\rm diss} \sim 0.38 - 0.41 $ foe.

\begin{figure}[!ht]
\begin{center}
\includegraphics[width=10cm]{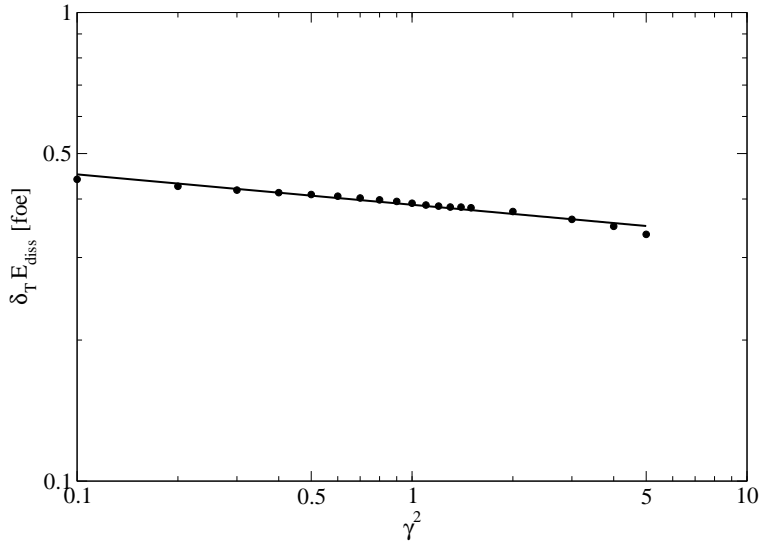}
\end{center}
\caption{Gain in dissociation energy of the shock, $\delta_{_T}
E_{\rm diss}$ (in foe), as a function of the strength of nuclear
electron capture, $\gamma^2$. The calculated points correspond to
standard parameters of the model ($\rho_{\rm{tr},10} = 43 $, $
m_{\omega}(0)= 1.7 $, $ T_0 = 2 $ MeV, $ \Delta_{\rm N}  = 2.5 $
MeV). The line represents a power-law fit, with exponent $m=
-0.065$. } \label{fig2}
\end{figure}

Although the previous discussion indicate a quite stable value
$\delta_{_T} E_{\rm diss} \sim 0.4 $ foe, we want to study the
robustness of such a result under reasonable variations of the
model parameters, compatible with present theoretical
uncertainties about the values of  $\rho_{\rm{tr},10}$,
$\Delta_{\rm N}$, $ m_{\omega}(0)$ and $ T_0 $. In
Table~\ref{tab2}, we show $\delta_{_T} E_{\rm diss}$ for different
values of $\gamma^2$: each column represents the case in which
only one of the parameters, $\rho_{\rm{tr},10}$ or $ \Delta_{\rm
N} $, is changed from its standard value to the value indicated.
In Figure~\ref{fig3}, instead, we show a contour plot for
$\delta_{_T} E_{\rm diss}$ (in foe) as a function of the two
parameters $ m_{\omega}(0)$ and $ T_0$, the other ones being fixed
at their standard values. The solid level lines are for $\gamma^2
= 1$, the dotted ones for $\gamma^2 = 0.1$ and the shaded area
indicates the physically meaningful range found in Paper~I for the
 thermal parameters of the symmetry energy.

\begin{table*}[!ht]
\begin{center}
\begin{tabular}{cccccc}
\hline\hline \ \ $\gamma^2 $ \ \ \ &  $ \Delta_{\rm N}  = 2 $ & $
\Delta_{\rm N}  = 3 $  &  $ \Delta_{\rm N}  = 4 $
&  $\rho_{\rm{tr},10} = 35 $  &  $\rho_{\rm{tr},10} = 55 $\\
\hline
0   & 0.45 & 0.45 & 0.45 & 0.43  & 0.46  \\
0.1 & 0.43 & 0.45 & 0.46 & 0.44  & 0.44  \\
0.5   & 0.40 & 0.42 & 0.43 & 0.42  & 0.38 \\
1   & 0.38 & 0.40 & 0.42 & 0.42  & 0.36  \\
2  & 0.37 & 0.38 & 0.40 & 0.40  & 0.34  \\
5   & 0.28 & 0.35 & 0.38 & 0.37  & 0.28  \\
 \hline\hline
\end{tabular}
\end{center}
\caption{Dependence of the results from the parameters of the
model. We show the gain in dissociation energy of the shock, $
\delta_{_T} E_{\rm diss}$ (in foe), for different values of the
strength of nuclear electron capture, $\gamma^2$. In each column
we change only the value of one parameter, either the excitation
energy of the GT resonance, $ \Delta_{\rm N} $ (in MeV), or the
trapping density, $\rho_{\rm{tr},10}$, while the other parameters
are the standard ones ($\rho_{\rm{tr},10} = 43 $, $ m_{\omega}(0)=
1.7 $, $ T_0 = 2 $ MeV, $ \Delta_{\rm N}  = 2.5 $ MeV). }

\label{tab2}
\end{table*}

\begin{figure}[!ht]
\begin{center}
\includegraphics[width=10 cm]{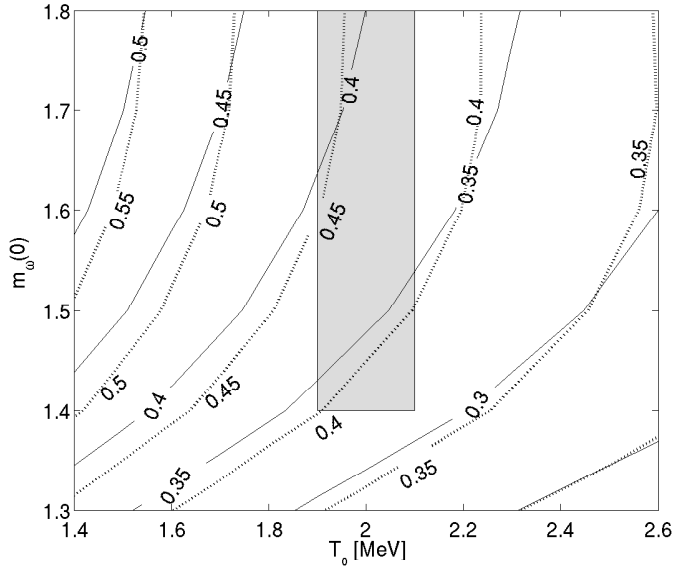}
\end{center}
\caption{Gain in dissociation energy of the shock, $\delta_{_T}
E_{\rm diss}$ (level lines labelled in foe), as a function of the
parameters $m_{\omega}(0) $ and $T_0$. The solid contour lines
correspond to $\gamma^2 = 1$, the dotted contour lines to
$\gamma^2 = 0.1$. The other parameters are the standard ones
($\rho_{\rm{tr},10} = 43 $, $ \Delta_{\rm N}  = 2.5 $ MeV). The
shaded area shows the physical range found in Ref.~\cite{DPBB} for
the parameters $m_{\omega}(0) $ and $T_0$. }\label{fig3}
\end{figure}

The results of Table~\ref{tab2} and Figure~\ref{fig3} show that,
under reasonable variations of the model parameters, the gain in
dissociation energy of the shock  changes only by about $ \pm 10\%
$, in the range $\delta_{_T} E_{\rm diss} \sim 0.35-0.45$ foe.
This proves the robustness of our conclusions: the temperature
dependence of the symmetry energy yields a systematic energy gain
(less dissipation of shock energy), whose order of magnitude is
$\delta_{_T} E_{\rm diss} \sim 0.4 $ foe\footnote{As in Paper~I,
we have set the volume  and  symmetry energy coefficients in the
BBAL EOS to the values $w_0 = -16.5$ MeV and $s(0) = 29.3$ MeV
respectively. Different values give a different energy gain, but
do not alter our general conclusions (for example, with $w_0 =
-16$ MeV and $s(0) = 31.3$ MeV  we find $\delta_{_T} E_{\rm diss}
\sim 0.3$ foe).}. In the concluding section, we will discuss the
relevance of such a result for the physics of supernova
explosions.

\section{\label{sec4}Conclusion}

In this article we have studied the effect of the temperature
dependence of the symmetry energy on the neutronization processes
occurring during gravitational collapse in a supernova explosion.
We have assumed for $E_{\rm sym}$ the $T$-dependence found in
Paper~I and later confirmed in Ref.~\cite{DLS}, first fixing the
parameters to their average values $ m_{\omega}(0)= 1.7 $, $ T_0 =
2 $ MeV, but later allowing them to vary in a reasonable physical
interval. We have followed the collapse with a one-zone model,
finding the collapse trajectories of the different thermodynamical
variables (temperature, entropy, particle abundances, lepton
fractions, mean heavy nucleus) and determining the equilibrium
lepton fraction after neutrino trapping, when the collapse becomes
adiabatic. We have implemented electron capture on both free
protons and nuclei with standard two-level transitions at finite
temperature. However,  to account for the present theoretical
uncertainties concerning electron capture rates in exotic nuclei,
we have multiplied the nuclear strength recently obtained in
Ref.~\cite{Lal} by a strength parameter $\gamma^2$:  variations in
a range $0.5 \lesssim \gamma^2 \lesssim 2$ around the presently
accepted value of $\gamma^2 = 1$ are not to be ruled out in the
future. Starting from the improved presupernova initial conditions
of Ref.~\cite{HLMW}, we have run the collapse simulation with and
without the $T$-dependence of $E_{\rm sym}$ implemented, thus
obtaining the ''thermal'' change in deleptonization, $\delta_{_T}
Y_{l,{\rm tr}}$. Then, we have studied the significance of this
thermal effect in terms of a quantity with more direct physical
meaning, the corresponding gain in dissociation energy of the
shock, $\delta_{_T} E_{\rm diss} \propto \delta_{_T} Y_{l,{\rm
tr}}^2 $. Finally, we have  tested the solidity of our results by
varying the standard parameters of the model ($\rho_{\rm{tr},10} =
43 $, $ m_{\omega}(0) =1.7$, $ T_0 =2 $ MeV, $ \Delta_{\rm N}  =
2.5$ MeV) within reasonable physical ranges, compatible with
present theoretical uncertainties.

The main conclusion of our investigation is that  the temperature
dependence of the symmetry energy systematically reduces the
neutronization of the core, namely it consistently increases the
equilibrium lepton fraction by a small constant amount,
$\delta_{_T} Y_{l,{\rm tr}} \simeq 0.006$, irrespective of the
value of the strength parameter $\gamma^2$. The corresponding gain
in shock dissociation energy, instead, decreases with increasing
nuclear strength, but \emph{very} slowly: when $\gamma^2=1$ is
divided or multiplied by two, $\delta_{_T} E_{\rm diss}$ varies
only by $\pm 4\% $ around its standard parameter value
$\delta_{_T} E_{\rm diss}|_{_{\gamma^2=1}} = 0.39$ foe. Moreover,
significant changes in the other one-zone model parameters
correspond to a quite small range $\delta_{_T} E_{\rm diss} \sim
0.35-0.45$ foe. This confirms the robustness of our results and
the presence of a systematic gain in shock dissociation energy of
order $\delta_{_T} E_{\rm diss} \sim 0.4$ foe, associated to the
temperature dependence of the symmetry energy.

Such an effect is obviously not a dramatic one, when one considers
that the total energy sapped from the shock by photo-dissociation
of nuclei is  larger by almost two orders of magnitude. Indeed,
even changing the nuclear strength by a  factor of ten through the
unblocking of GT transitions does not qualitatively alter the
final outcome of the failed explosion, at least in one-dimensional
simulations \cite{Hal1}. Actually, recent developments of
three-dimensional simulations of core-collapse supernovae indicate
that the roles of neutrinos, fluid instabilities, rotation and
magnetic fields are probably critical to obtain successful
explosions \cite{WJ}. However, when compared to the typical
kinetic energies of a supernova explosion, $K_{\rm expl}$, which
are imparted by the shock wave to the ejecta, a gain in shock
energy of $\delta_{_T} E_{\rm diss} \sim 0.4$ foe is not
negligible (for SN 1987A, observation gave $K_{\rm expl} \sim 1 $
foe \cite{BP}). Moreover, $\delta_{_T} E_{\rm diss}$ is two orders
of magnitude larger than the total electromagnetic output
\cite{Be}. On general grounds, since both the explosion energy
$K_{\rm expl}$ and the much smaller electromagnetic output have
small values resulting from differences of very large quantities
(gravitational energy, initial post-bounce shock energy, neutrino
losses, nuclear photo-dissociation), it follows that the explosion
observables  can be sensitive to subtle microphysical features. In
particular, systematic nuclear effects can be of particular
importance, as noted also in the conclusions of Ref.~\cite{Hal2}.

The numerical results of our one-zone collapse simulation are
significant for their order of magnitude, not their precise values
which are limited by the oversimplified zero-dimensional approach.
In our opinion, their robustness under variations of the model
parameters justifies further investigation in detailed
one-dimensional numerical codes. This is not a straightforward
task, since not only the reaction Q-values, but the whole EOS
describing dense hot matter is affected by the $T$-dependence of
$E_{\rm sym}$. In particular, this is lengthy to implement within
the Lattimer and Swesty EOS \cite{LS}, currently used in realistic
supernova codes, since changing any of its nuclear input
parameters (in the present case introducing $ m_{\omega}(0)$ and
$T_0$ to parameterize $ m^{\star} = m^{\star}(T)$  and thence the
temperature dependence of the symmetry energy) requires
re-calculating the phase and Maxwell construction boundaries
before generating the final table of the EOS. We are presently
working along these lines \cite{Fal}, with interesting preliminary
results from one-dimensional simulations with the simpler BBAL
EOS, where the temperature dependence of the symmetry energy can
be implemented analytically.

{\bf Acknowledgements}

This work was (partially) supported by \emph{CompStar}, an ESF
Research
Networking Programme (www.esf.org/compstar). \\
P.M.P. dedicates this work to Gerry Brown, who originally
suggested both the increase with temperature of the symmetry
energy and its possible relevance for supernova explosions.

\end{document}